# 3D density structure and geological evolution of Stromboli volcano (Aeolian Islands, Italy) inferred from land-based and sea-surface gravity data

Niklas Linde[1*], Ludovic Baron[1], Tullio Ricci[2], Anthony Finizola[3], André Revil[4,5], Filippo Muccini[6], Luca Cocchi[6], and Cosmo Carmisciano[6]

[1]Applied and Environmental Geophysics Group, Institute of Earth Sciences, Faculty of Geosciences and the Environment, University of Lausanne, Switzerland;

[2]Istituto Nazionale di Geofisica e Vulcanologia, Rome, Italy;

[3]Laboratoire GéoSciences Réunion, Université de la Réunion, Institut de Physique du Globe de Paris, Sorbonne Paris-Cité, CNRS UMR7154, Saint-Denis, La Réunion, France;

[4]Colorado School of Mines, Department of Geophysics, Golden CO, USA;

[5] ISTerre, CNRS, UMR CNRS 5275, Université de Savoie, Le Bourget du Lac, France;

[6] Istituto Nazionale di Geofisica e Vulcanologia, Porto Venere, Italy.

* Corresponding author: Niklas Linde

University of Lausanne

Géopolis - bureau 3779

CH-1015 Lausanne

Email : Niklas.Linde@unil.ch

Phone :    +41 21 692 4401

Fax :    +41 21 692 44 05

**Abstract**. We present the first density model of Stromboli volcano (Aeolian Islands, Italy) obtained by simultaneously inverting land-based (543) and sea-surface (327) relative gravity data. Modern positioning technology, a 1 × 1 m digital elevation model, and a 15 × 15 m bathymetric model made it possible to obtain a detailed 3-D density model through an iteratively reweighted smoothness-constrained least-squares inversion that explained the land-based gravity data to 0.09 mGal and the sea-surface data to 5 mGal. Our inverse formulation avoids introducing any assumptions about density magnitudes. At 125 m depth from the land surface, the inferred mean density of the island is 2380 kg m$^{-3}$, with corresponding 2.5 and 97.5 percentiles of 2200 and 2530 kg m$^{-3}$. This density range covers the rock densities of new and previously published samples of Paleostromboli I, Vancori, Neostromboli and San Bartolo lava flows. High-density anomalies in the central and southern part of the island can be related to two main degassing faults crossing the island (N41 and N64) that are interpreted as preferential regions of dyke intrusions. In addition, two low-density anomalies are found in the northeastern part and in the summit area of the island. These anomalies seem to be geographically related with past paroxysmal explosive phreato-magmatic events that have played important roles in the evolution of Stromboli Island by forming the Scari caldera and the Neostromboli crater, respectively.

# 1. Introduction

The island of Stromboli (surface area 12.6 km$^2$) in the Aeolian archipelago (Southern Italy, north of Sicily) is part of a volcanic arc that developed along a NE-SW regional extensional fault system. It rises 2400-2700 m above the sea floor and peaks at 924 m above sea level. Stromboli is characterized by a complex geological structure caused by the interplay of six distinct eruptive epochs and important vertical caldera-type and lateral collapses. These collapses have largely conditioned the deposition of younger products by providing topographic lows, but also barriers to lava flows. They have also played key roles in ending eruptive epochs (Hornig-Kjarsgaard et al., 1993; Pasquarè et al., 1993; Tibaldi, 2010; Francalanci et al., 2013). In contrast to the many detailed geological studies, very few attempts have been made to image the internal 3-D structure of Stromboli using geophysical methods.

Bossolasco (1943) performed a land-based magnetic survey on Stromboli, while Okuma et al. (2009) presented the first 3-D model by inverting airborne magnetic data. They found an important magnetization low below the summit craters that they explained by demagnetization caused by the heat of conduits and hydrothermal activity, as well as accumulation of less magnetic pyroclastic rocks. The magnetic highs are located in areas exposed by basaltic-andesite to andesite lavas. Bonasia and Yokoyama (1972) and Bonasia et al. (1973) presented gravity data from Stromboli. They found a Bouguer anomaly low in the central part of the island using 37 relative gravity measurements with a vertical positioning accuracy of ± 3 m. Their interpretation of a corresponding density-low in the summit area is questionable as no terrain or bathymetric corrections were carried out (see discussion in Okuma et al., 2009). Indeed, topographic and bathymetric effects on volcanic islands are extremely important and will, if left unaccounted, mask any information about density variations. It is thus very likely that the negative Bouguer anomaly inferred by Bonasia et al. (1973) in the central part of the island is mainly caused by unaccounted topography. Furthermore, their positive Bouguer anomalies to the northeast and southwest are likely due to unaccounted effects of shallow platforms located below the sea level (Gabbianelli et al., 1993). This implies that no reliable information exists to date about the density structure of Stromboli.

Three-dimensional inversions of gravity data from volcanic islands are rather common. For instance, inversions have been performed using gravity data acquired over the Canary Islands (e.g., Montesinos et al., 2006; 2011), the Azores (e.g., Represas et al., 2012), and at La Soufrière volcano (Coutant et al., 2012). These studies typically relied on 93 to 365 land-based gravity data (and sometimes sea-surface data; e.g., Montesinos et al., 2006). They definitively helped to improve the understanding of structural settings and their control on volcanic activity.

Gravity inversions are typically performed using finely discretized models and least-square methods that seek smooth property variations (e.g., Coutant et al., 2012) or methods that seek the appropriate location, shape and volume of anomalies with predefined density contrasts (Camacho et al., 2000; Montesinos et al., 2005). The first category refers to linear inverse problems that are easily solved, but the resulting models have smoothly varying property variations that make the identification of geological contacts difficult. The second category leads to more time-consuming nonlinear inverse problems, but allows resolving the volume of anomalous bodies provided that the appropriate density contrasts are known. Previous gravity studies on volcanic islands (see citations above) suggest that the quality of the density models are not only dependent on the coverage and quality of the gravity data, but that they are also strongly affected by (1) the resolution and precision of the topographic and bathymetric models and (2) how this information is included in the inversion.

We present results from the first detailed land-based gravity survey on Stromboli. A total of 543 gravity stations were complemented with a subset of 327 sea-surface gravity data. The data were inverted in 3-D to better understand the geological structure at depth and its control on the hydrothermal system. The inversion incorporated a high-resolution and precise digital elevation model (DEM) including the bathymetry. The resulting density model was interpreted in the light of previous geophysical studies and present-day geological understanding of this volcanic edifice.

## 2. Geological setting

The edification of the emerged part of Stromboli can be subdivided in the following six main epochs (Francalanci et al., 2013), shown in Figure 1:

(1) Epoch 1: (Paleostromboli I period: from 85 to 75 ka). This period is mainly associated with massive to blocky lava flows and pyroclastic products. Paleostromboli I ended with the formation of the Paleostromboli I caldera (see "PST I" in Fig. 1).

(2) Epoch 2: (Paleostromboli II period: from 67 to 56 ka). This second epoch is characterized by massive to blocky lava flows interbedded with scoriaceous deposits and ended with the Paleostromboli II caldera that can be evidenced in Vallone di Rina.

(3) Epoch 3: (Paleostromboli III period and Scari Units: from 56 to 34 ka). The Paleostromboli III period is particularly developed in the Vallone di Rina. Sub-period 3a displays mainly pyroclastic products with lava flows alternating with scoriaceous beds in the upper part of the geological succession and ends with a caldera formation. Sub-period 3b is mainly associated with lava flows. This period ended with the hydromagmatic Scari Unit deposits, located in the northeastern part of the island. Nappi et al. (1999) suggested its eruptive center from sector of provenance of ballistic ejecta (see Fig. 1). No caldera boundary has been evidenced in this area, probably due to its refilling by younger products. Epoch 3 ended with the formation of the Paleostromboli III caldera ("PST III" in Fig. 1).

(4) Epoch 4: (Vancori Period: from 26 to 13 ka). The Vancori period is characterized by successions of lava flows and is subdivided into three sub-periods 4a, 4b, 4c, separated by a caldera formation, a quiescence period and a sector collapse.

(5) Epoch 5: (Neostromboli Period: from 13 to 4 ka). The Neostromboli period is essentially characterized by lava flows and scoriaceous deposits and it is subdivided into three sub-periods 5a, 5b, 5c, separated by sector collapse, and two strong hydromagmatic eruptions. These eruptions associated with pyroclastic and pumice deposits (Punta Labronzo deposits) were responsible for the formation of the Neostromboli crater (Fig. 1).

(6) Epoch 6: (Pizzo and Present-day activity: since 2 ka). This last period is subdivided into 3 sub-periods. Sub-period 6a is associated with the pyroclastic successions related to the Pizzo activity, lava flows, such as, San Bartolo (Fig.1) and it ends with the Rina Grande sector collapse (Fig. 1). Sub-period 6b is associated with scoriaceous and pumiceous products of the Present-day activity, massive lava flows, and the Sciara del Fuoco sector collapse (Fig. 1). Sub-period 6c began after this last major collapse (1631-1730 AD) and is characterized by scoriaceous (pumiceous) and lava flow deposits related to Present-day activity in the Sciara del Fuoco area, and to reworked scoriaceous product in the Rina Grande area. The most recent effusive eruption of Stromboli took place in 2007 from February 27 to April 2. This eruptive event was characterized by persistent lava flows along Sciara del Fuoco and by a paroxysmal explosion on March 15.

During these six main epochs of activity, lava flows can be considered as the main eruptive dynamics of the emerged part of the Stromboli edifice.

## 3. Method

### 3.1. Forward modeling

The least-square smoothness-constrained gravity inverse problem is linear and easy to solve, but inversion results can be severely affected by inaccurate forward modeling. Our 3-D forward model was thus designed to accommodate precise positioning, a high-quality DEM with a resolution of 1 × 1 m (Marsella and Scifoni, private communication) covering the aerial part of the island and a bathymetric model with a resolution of 15 × 15 m (Casalbore et al., 2011).

The modeling domain was discretized by rectangular parallelepipeds. The vertical component of the gravity response of each parallelepiped was calculated using the analytical solution of Banerjee and Das Gupta (1977). To accurately account for the bathymetry and its effect on the gravity data, the discretized modeling domain had a lateral extent exceeding 20 km. In addition, external forward model cells were extended $10^6$ m to the sides to avoid boundary effects. A 10 × 10 m resolution model was derived from the mean values of 10 × 10 m blocks of the DEM and by interpolation of the bathymetric model. The gravity forward model used this 10 × 10 m resolution model to calculate the integrated response of larger inversion cells when these cells intersected the land surface or the sea floor. The forward model discretization was further refined to 1 × 1 m for inversion cells centered within 100 m of a given measurement location.

### 3.2. Inverse modeling

The smoothness-constrained least-squares inverse problem consists of solving the following system of equations in a least-squares sense (e.g., Coutant et al., 2012):

$$\begin{bmatrix} \mathbf{C}_d^{-0.5}\mathbf{F} \\ \lambda \mathbf{W}_m \end{bmatrix} [\mathbf{m}] = \begin{bmatrix} \mathbf{C}_d^{-0.5}\mathbf{d}' \\ \mathbf{0} \end{bmatrix}, \quad (1)$$

where $\mathbf{C}_d$ is the data covariance matrix describing the data errors and it is here assumed to be adequately represented by uncorrelated Gaussian data errors of a known standard deviation (i.e., 0.1 mGal for the land-based data and 5 mGal for the sea-surface data), $\mathbf{F}$ is the forward kernel that provides the gravity responsefor a unit density with respect to a base station, $\mathbf{d}'$ is the processed relative gravity data, $\mathbf{W}_m$ is the model regularization operator (a discretized gradient operator is used in this study), $\lambda$ is the regularization weight that determines the weight given to the model regularization term, and $\mathbf{m}$ is the resulting model.

Note that the reference used for calculating **d'** and **F** can be different for different data sources. Furthermore, the reference does not necessarily have to refer to a given reference point, but can also, for instance, be the average response of several data points. In the following, we will use the calculated sea-surface response with respect to the average response of all sea-surface data, as there was no near-by base station for these data. The inverse problem was solved with LSQR (Paige and Saunders, 1982) by varying the regularization weight $\lambda$ by trial-and-error until the data residuals were similar to the assumed standard deviation of the data errors.

To resolve sharper transitions in model properties, we carried out additional iterations using an iteratively reweighted least-squares procedure that minimized a perturbed $l_1$ model norm following Farquharson (2008). To avoid being overly sensitive to data outliers (the case when assuming a Gaussian error distributions), we also applied an iterative reweighting of the data residuals to imply a more robust perturbed $l_1$ data norm. The resulting system of equations to solve in a least-squares sense at the $p+1$ iteration is:

$$\begin{bmatrix} \mathbf{R}_{d,p}\mathbf{C}_d^{-0.5}\mathbf{F} \\ \lambda \mathbf{R}_{m,p}\mathbf{W}_m \end{bmatrix} [\mathbf{m}_{p+1}] = \begin{bmatrix} \mathbf{R}_{d,p}\mathbf{C}_d^{-0.5}\mathbf{d}' \\ \mathbf{0} \end{bmatrix}, \quad (2)$$

where $\mathbf{R}_{d,p}$ and $\mathbf{R}_{m,p}$ are diagonal reweighting matrices. The diagonals are defined as

$$\mathbf{r}_{d,p} = \left[ \left( \mathbf{C}_d^{-0.5}(\mathbf{d}' - \mathbf{d}_p^{pred}) \right)^2 + \left( \gamma_{d,p} \right)^2 \right]^{-0.5}, \quad (3)$$

$$\mathbf{r}_{m,p} = \left[ \left( \mathbf{W}_m \mathbf{m}_p \right)^2 + \left( \gamma_{m,p} \right)^2 \right]^{-0.5}, \quad (4)$$

with $\gamma_{d,p}$ and $\gamma_{m,p}$ being small scalar values defined similarly to Rosas Carbajal et al. (2012) as $0.5 \times \mu(\mathbf{W}_m \mathbf{m}_p)$ and $0.5 \times \mu(\mathbf{C}_d^{-0.5}(\mathbf{d}' - \mathbf{d}_p^{pred}))$, respectively, with $\mu(-)$ denoting the mean value and $\mathbf{d}_p^{pred}$ the simulated model response of model $\mathbf{m}_p$ at the previous iteration. A classical least-squares inverse problem using equation (1) was solved at the first iteration, $p=1$, while equation (2) was used in subsequent iterations, $p>1$.

The inverse model parameterization was based on cubes with side-lengths of 50 m, which is in agreement with Coutant et al. (2012). The cubes extended from the land surface down to 500 m depth below sea level from which parallelepipeds with a vertical extent of 5500 m (horizontal dimensions of 50 m) were extended down to 6000 m depth. A total of 179,172 inversion cells were used. As explained above, the forward operator **F** accounted for the topography at a resolution of 10 m with local refinement to 1 m in the vicinity of the gravity stations.

The inverse formulation described above in equations (1-4) provides a density model for a data set that is solely affected by the density distribution of the solid earth in the surroundings of the survey area. Before we can use this formulation to invert the gravity data acquired at Stromboli, we must thus remove all unrelated effects from the data.

*3.3. Data acquisition*

The land-based gravimetric survey was designed to achieve a suitable data coverage given logistic constraints. Extremely dense vegetation in the lower part of the island (0-500 m elevation) limited access to a few pre-existing paths and to profiles that had been cleared out for other field campaigns or for this survey. The region above the vegetation limit is generally accessible, except for some areas where the risks associated with rock falls, landslides or volcanic activity are too important. These constraints resulted in data gaps, notably in the Sciara del Fuoco (see Fig. 1) and in the immediate region surrounding the active vents.

The first part of the gravity survey was carried out January 12-24, 2012. Based on initial inversion results and data coverage, the survey was completed in the period of September 22-30, 2012. The resulting station coverage (see Fig. 2) is characterized by close station spacing along profiles (50-80 m), but sometimes large (km scale) separation between profiles due to the logistic constraints described above. A relative gravimeter (CG-5, Scintrex) was used to perform measurements at 556 unique locations, with 56 of them being measured repeatedly to determine the instrument drift. The gravity data were acquired using a measurement frequency of 5 Hz and stacking during 30 s. This sequence was repeated five times at each station and the median value was recovered for further analysis. The data acquisitions were reinitiated if the noise level increased (typically more than ten-fold) due to volcanic activity (see Carbone et al., 2012) or when the magnitude of the tilt was above 10 arcseconds at the end of the measurements. The station locations were determined using a differential ground positioning system (DGPS). A Rover 1200 by Leica was used with a measurement time of approximately 2 minutes. Most of the GPS data were post-processed using a permanent GPS station (SVIN) installed by Istituto Nazionale di Geofisica e Vulcanologia (http://ring.gm.ingv.it; Selvaggi et al., 2006), for monitoring purposes, close to the civil protection center (COA) on the island. Our own base station was used for DGPS processing when the monitoring station was out of order. The DGPS processing was carried out with the publicly available software library RTKLIB (http://www.rtklib.com/) using the IGS broadcast and precise orbit (Dow et al., 2009).

We used a small subset (327 data points with a spacing of ~10 data points per km$^2$) of the sea-surface gravity data acquired during the 2010 PANSTR10 scientific cruise onboard R/V Urania using an AirSea MicroG gravity meter (Bortoluzzi et al., 2010) and DGPS for positioning. Inside the ship, the gravimeter was placed on a stabilized platform with a 4-minute period. The recorded gravity data were the outcomes of a Blackman filter that averaged the instrument response over 4 minutes. These are the data used for subsequent data processing.

*3.4. Data processing*

Before being able to apply our inversion algorithm, we had to correct our recorded data for tidal effects, instrument drifts, latitude effects, free-air, the gravitational acceleration of the surrounding sea and regional effects. For the sea surface data, it was also necessary to account for the ship trajectory. These corrections are described in the following.

The data were first corrected for tidal effects before applying an instrument drift correction. The daily instrumental drift was computed by repeating the measurements at known locations in the beginning and in the end of the day and assuming that the drift was linear with time. For the land-based data, the applied drift correction was 0.028 mGal on average and 0.134 mGal at the most. To further evaluate the data quality of the land-based data, the resulting data were compared at 56 stations that were measured at different times. The discrepancies between repeated measurements (after tide and drift corrections) were 0.055 mGal on average and the largest difference was 0.125 mGal. For the actual gravity measurements, the standard deviation of the recorded average response was 0.003 mGal on average and 0.023 mGal at the most. It appears thus safe to assume a total standard deviation of the accumulated data error of 0.1 mGal on all land-based data, which is in agreement with Represas et al. (2012).

A free-air correction was applied to correct the tide- and drift-corrected data for the variation of standard gravity with altitude (e.g., Lowrie, 2007). The resulting data were corrected for latitude using the normal gravity formula defined for the Geodetic Reference System (GRS80) (e.g., Lowrie, 2007). The sea effect for each datum was calculated using our forward model (section 3.2.).

The sea-surface data were adversely affected by the ship trajectory. One important effect associated with a moving acquisition platform (i.e., the ship) is the Eötvos acceleration that leads to a decrease of the measured vertical acceleration of gravity when the ship moves eastwards and to increases when the ship moves westwards. This effect is caused by changes in the centrifugal acceleration associated with the rotation of the Earth. It is possible to minimize the Eötvos acceleration by using a heading that is predominantly N-S, but the effect is very important when the ship moves to the east or west, which was often the case in the vicinity of the island (± 60 mGal in the field data). Another effect relates to the centrifugal acceleration exerted by ship turns. This effect was predicted to be on the order of -20 mGal using calculations based on Swain (1996). An important consequence of the stabilized platform (4-minute period) and the corresponding time averaging of the data is that the imprints of the Eötvos acceleration and ship turns remain for long times in the recorded data. For a typical boat speed of 5 m s$^{-1}$, the data averaging is made over 1.2 km. Despite the precise positioning offered by DGPS, it is most difficult to accurately predict these effects, as the detailed response of the stabilized platform is unknown. This leads to spatially correlated errors in the vicinity of the island that are on the order of several mGal.

The sea-surface data were used to establish a regional trend model that was subsequently removed from both the sea-surface and the land-based data. The linear regional trend predicted increasing values towards the north at a rate of 1.74 mGal/km and increasing values to the east with a rate of 0.16 mGal/km. After all the corrections had been carried out, the land-surface data were referred to one reference location, while the sea-surface data were referred to the mean of all the sea-surface data. These corrected data were then used as **d**' in our inverse formulation (equations (1-4)) together with a forward kernel **F** that appropriately accounted for the different references used for the land-based and the sea-surface data. Note that no assumptions were made about the density values of the solid earth when carrying out these corrections. In the following, we assume that **d**' are only sensitive to the shape and density distribution of Stromboli.

## 4. Results

### 4.1. Sea effect and local Bouguer anomaly

The vertical gravitational acceleration of the sea (referred to as sea effect in the following) was calculated using a seawater density of 1030 kg m$^{-3}$ (see Fig. 2). The magnitude and variation of the sea effect across the island (between 15-29 mGal) illustrate that an inaccurate or too coarse bathymetric model makes any detailed gravity analysis most problematic. The sea effect is even more important for the sea-surface data acquired above the deepest sea (up to 63 mGal), but the largest effects on the island are found in the summit area.

The accuracy and resolution of the DEM describing the island topography are most important (e.g., Coutant et al. (2012) estimated at La Soufrière that the error due to an imperfect DEM (2 m error in elevation at a resolution of 5 m) was below 0.5 mGal). To evaluate the accuracy of the 1 × 1 m DEM, we compared our land-based DGPS altitudes with the closest node point of the DEM. This comparison excluded 13 of the 556 gravity stations that were not further used in the gravity inversions. Points were excluded when (1) the gravity reading was clearly identified as an outlier with respect to neighboring stations or when (2) the discrepancy between the DEM and DGPS was above 2 m. The standard deviation between the DEM and the DGPS for the remaining 543 stations was 0.25 m, which suggests an excellent quality both in terms of the DEM and the DGPS.

Distinct calculations of the Bouguer anomalies were done for the land-based and the sea-surface data. In each case, local Bouguer anomalies were calculated by differencing **d**' and the forward response for a uniform density of the solid earth. The uniform densities were chosen as 2100 kg m$^{-3}$ for the land-based and 2600 kg m$^{-3}$ for the sea-surface data. These values were chosen to provide, in each case, Bouguer anomalies that were only weakly correlated with altitude or bathymetry. These Bouguer anomalies were calculated to represent the data and were in no way used for the subsequent inversion (i.e., the chosen densities have no influence whatsoever on the final inversion results).

Variations of the land-based local Bouguer anomalies (Fig. 3a) are rather small (-5 to 2 mGal; -10 to 2 mGal without removing the regional trend estimated from the sea-surface data). Negative anomalies are found over the Pizzo and South of Vancori, as well as in a zone east-northeast of the Pizzo (see locations in Fig. 1). Positive anomalies are found in the Fossetta, in the Sciara del Fuoco west of the active vents, in the region above Punta Labronzo, and in the vicinity of Piscità. The local sea-surface Bouguer anomalies (Fig. 3b) indicate that the sea-surface data are strongly affected by correlated data errors. Indeed, the predominant N-S acquisition profiles display rather smooth variations, while larger differences are seen when comparing parallel profiles. This suggests that no well-resolved information about the density structure can be resolved from the sea-surface data at this scale. A more advanced data processing and filtering could probably improve the situation, but the data errors appear almost 50 times higher than for the land-based data (Fig. 4). Sea-bottom gravity (Berrino et al., 2008) or acquisitions with a ship that moves much slower would solve many of these problems, but the data acquisitions would be much more time consuming.

*4.2. Inversion results*

The inversion result presented herein corresponds to the model obtained after one IRLS iteration with $\lambda = 0.002$. The data residuals have a standard deviation of 0.092 mGal for the land-based data and 5 mGal for the sea-surface data, while the corresponding mean deviations are 0.071 and 3.9 mGal. The data misfit for the land-based data is remarkably low compared to previous investigations on volcanic islands. Applications of the growing-bodies inversion based on Camacho et al. (2000) or similar inversion algorithms have previously resulted in data residuals on the order of 0.58 to 1.77 mGal (Araña et al., 2000; Montesinos et al., 2006; Montesinos et al., 2011; Represas et al., 2012). The large residuals in previous investigations are probably not related to the gravity data themselves (i.e., the standard deviation of the data are often close to 0.1 mGal; e.g., Represas et al., 2012), but appears to be related to coarse bathymetric (1' grid resolution in Represas et al., 2012) and DEM models (20 m resolution and 2 m error in Represas et al., 2012); terrain corrections based on constant density values; or generally coarse model discretization (900 m and coarser in Montesinos et al., 2011). For an inversion strategy similar to our own, Coutant et al. (2012) explained the gravity data at La Soufrière to 0.79 mGal. This higher data misfit is likely caused by a less precise DEM (vertical accuracy between 2 and 5 m) and bathymetric model than the ones available at Stromboli.

Figure 4a displays a W-E vertical cut through the model, while Figure 4c displays a S-N cut at the locations outlined in Figure 4b. The W-E profile indicates that the main dense anomaly (maximum value 2570 kg m$^{-3}$) is located SW of the active craters (see locations in Fig. 1). The S-N profile indicates that the ridge south of Rina Grande has a low density (minimum value 2140 kg m$^{-3}$), that the upper part of Rina Grande is dense (maximum value 2710 kg m$^{-3}$), while the Pizzo is characterized by low densities (minimum value 2030 kg m$^{-3}$).

A horizontal slice through the model at sea level (Fig. 5a) displays a major high-density anomaly that covers the central part of the island and continues towards the southwest (maximum value 2480 kg m$^{-3}$), while a low-density region is located towards the northeast (minimum value 2250 kg m$^{-3}$). At an elevation of 500 m (Fig. 5b), the highest densities are found in the surroundings of the Neostromboli crater (see Fig. 1) (maximum value 2560 kg m$^{-3}$). At 800 m (Fig. 5c), the Pizzo ridge (see Fig. 1) is found to have a low density (minimum value 2060 kg m$^{-3}$), while the Vancori ridge (see Fig. 1) has a high density (maximum value 2650 kg m$^{-3}$). Figure 5d is a slice of the inversion model at a depth of 125 m with respect to the land surface (i.e., parallel to the topography). This representation is favored as it allows comparing the results across the island in one unique image at a resolution that is approximately constant. It highlights a high-density region surrounding the Neostromboli crater (maximum value 2620 kg m$^{-3}$) and that important zones of high densities, at lower elevations, are only seen towards the southwest. The most prominent low-density anomalies are those associated with the Pizzo ridge (minimum value 2090 kg m$^{-3}$) and a large zone towards the northeast (minimum value 2130 kg m$^{-3}$).

To evaluate the influence of the regional trend model on the inversion results, we performed an additional inversion without removing the trend from the gravity data. The resulting model of the island was very similar (not shown) to the one presented here as any large-scale trends in the data were effectively accommodated by smaller variations in density for inversion blocks located offshore. This suggests that the inversion results in the interior part of Stromboli are robust with respect to large-scale trends in the data.

In this study, the sea-surface data served mainly to establish the regional trend model and to assure that the predicted responses offshore were largely in agreement with the available sea-surface data. An error analysis of the corrected sea-surface data in the vicinity of Stromboli suggests that these errors were not only quite large, but also highly correlated (see section 3.3). When including additional sea-surface data in the immediate vicinity of the island (not shown in Fig. 3b), artificial gravity gradients appeared that could only be explained by unrealistic density variations. To avoid affecting the density estimates on the island, we decided to follow a conservative use of these data as outlined above (i.e., an assumed standard deviation of 5 mGal and ignoring the sea-surface data acquired close to the coast). A different and more advanced processing and modeling framework would be needed to appropriately include the sea-surface data in the vicinity of the island, but this is outside the scope of the present contribution.

## 5. Discussion

### 5.1. Comparison with density measurements on rock samples

Density variations between different geological units are primarily determined by porosity and water content, while effects related to the mineralogical composition are only expected to have a relatively minor influence. Apuani et al. (2005) investigated the densities of 13 lava samples from Stromboli (Fig. 6). They found grain densities in the range of 2620-2920 kg m$^{-3}$, while the ranges of bulk densities for unsaturated and saturated conditions were 2270-2580 kg m$^{-3}$ and 2340-2610 kg m$^{-3}$, respectively. The estimated porosities of these samples ranged from 3 to 20%. Apuani et al. (2005) also analyzed recent deposits in the Sciara del Fuoco consisting of sand and gravel fractions and estimated grain densities in the range of 2910-3080 kg m$^{-3}$. Using unsaturated bulk densities of 1320-1610 kg m$^{-3}$, they estimated porosity to be in the range of 40-55%.

Eight representative rock samples acquired during our field experiments were analyzed for grain, dry and saturated bulk density and porosity (see Table 1 for the values and Figure 6 for the sampling locations). The dried rock samples (0.7-2.0 kg) were first weighted using a precision balance (MS32001LE by Mettler-Toledo; 0.1 g for weights up to 32 kg). The rock samples were then fixed to a thin copper wire and immersed in a water bath. Immediately after immersion, the weight increase of the water corresponds to the weight of water displaced by the rock sample (and the copper wire). The rock samples were sprayed to decrease water imbibition, but some water uptake occurred over time. The measured weight at the time of immersion was unstable due to minor perturbations during this process. The weight at the time of immersion was hence estimated by using a high measurement frequency (23 Hz) and by fitting the decreasing weight of the water bath with a polynomial function. The large sample size, the high measurement frequency, and the precise balance allowed estimates of dry bulk density that were typically precise within 10 kg m$^{-3}$. The Rina Grande reworked sands were analyzed by weighting 1 liter of the sample. The weight was largely dependent on the packing method used to fill the tube and the associated error is estimated to be rather large (50 kg m$^{-3}$).

The grain densities were obtained by crushing parts of the rock samples (55-83 g). The density of water was first estimated using the same balance by filling 100 ml of water in a beaker. The weights of the dry crushed samples were then measured in the empty beaker. To estimate the corresponding sample volumes, the samples were saturated with water by careful stirring to avoid entrapped air. The beaker was then filled with the necessary amount of water to fill it up to 100 ml, which allowed estimating the volume of the crushed samples and, hence, the dry bulk density. From the estimated dry bulk and grain densities it is straightforward to derive the saturated bulk densities and the porosities. Our Neostromboli lava sample (sample 5) with a dry bulk density of 2230 kg m$^{-3}$ was in close agreement with two corresponding samples at Ginostra by Apuani et al. (2005) of 2270 kg m$^{-3}$. A good agreement was also found between our Paleostromboli I (sample 1) with a dry bulk density of 2310 kg m$^{-3}$ and the one of Apuani et al. (2005) at Malpasso of 2430 kg m$^{-3}$. No comparison was possible for the Vancori unit as the samples were collected at very different locations and the variability within this stratigraphic unit is very high.

The inversion result displays a density range that is in agreement with the bulk densities reported by Apuani et al. (2005) and our own density measurements on the lava samples. The mean density at a depth of 125 m from the land surface is 2380 kg m$^{-3}$ (see Fig. 5d). The corresponding 2.5 and 97.5 percentiles are 2200 kg m$^{-3}$ and 2530 kg m$^{-3}$, respectively. Our four lava flow samples have dry bulk densities that are located within this 2.5-97.5 percentile density range (see Fig. 6). The other four samples (altered (4), explosive (2), vesiculated (6) and reworked (8)) have much lower dry bulk densities indicating that such dry samples do not constitute important volumes at depth. In terms of water saturated bulk density, the altered (4) and explosive (2) samples reach densities that approach the 2.5 percentile of the density model. The good agreement between the density range for dry and water saturated lava flow (our samples (1, 3, 5 and 7) and Apuani et al., 2005 (all samples)) and the density range of the gravity model suggests that dry or saturated lava flows constitute the main volume of Stromboli.

*5.2. Geological interpretation*

One striking feature of the density model is the low-density anomaly (2100-2250 kg m$^{-3}$) on the Pizzo ridge with a possible extension towards the east within the Neostromboli crater and the high densities on the surrounding crater ridges (2500-2550 kg m$^{-3}$; see Fig. 7). The low densities of the pyroclastics and scoria deposits at Pizzo (in comparison to the dense lava flow of Vancori Unit) are easily explained by their high porosities and permeabilities, meaning that they effectively drain water, thereby implying not only high porosities, but also low saturation levels. Furthermore, it is expected that the paroxysmal activity in the summit area has deposited thick accumulations of highly vesiculated products in the depression areas, that is, within the Neostromboli crater. The surrounding high-density bodies are interpreted as being related to the past feeding system of the volcano, in which basaltic magma has risen and cooled very slowly, thereby forming very dense materials. Represas et al. (2012) performed density measurements on dyke samples at Maio island (Cape Verde) and found bulk densities ranging between 2690 and 3040 kg m$^{-3}$. The high-density region appears to continue towards the southwest, which is in agreement with the main region of dyke intrusions on Stromboli (Tibaldi et al., 2009). Security concerns limited the gravity measurements in the Sciara del Fuoco to a short profile west of the active craters. This area is clearly very dense and could be associated to old magmatic intrusions.

High-densities are found in the Rina Grande sector, which is generally thought to be part of the active hydrothermal system as evidenced by high electrical conductivities (e.g., Revil et al., 2011) and $CO_2$ emissions (Carapezza et al., 2009). In the northeastern part of the island, up to the village of San Vincenzo and inland, there is an important low-density anomaly of kilometric scale. The maximum of the high-density anomaly in the Rina Grande sector and the minimum of the low-density anomaly in the northeastern part of the island are aligned with the N64° fault defined with soil gas measurements by Finizola et al. (2002) (see Fig. 7). Both these anomalies could be interpreted in terms of areas of dyke intrusions along this structural boundary. In the Rina Grande area, dyke intrusions did not reach an eruptive stage. On Stromboli, based on a magma degassing budget, only ~1% of the degassing magma reaches the surface (Allard et al., 1994). In other words, dyke intrusions play an important role in such a volcanic edifice. For the low-density anomaly located in the northeastern flank of the island, the same dyke intrusion can be considered, but this time, the intrusion reached the surface. This part of the island is characterized by a special formation named the Scari Unit (see Fig. 1), which is composed of thick layers of tuffs (Keller et al., 1993; Francalanci et al., 2013). The upper part of the Scari unit is associated with a caldera formation suggested by Nappi et al. (1999) and located based on the direction of the ballistics fallen in the phreato-magmatic deposits of the Upper Scari Unit (see Fig. 1). This angle of location fits very well with a caldera centered on the low-density anomaly found in the area (Figs. 7, 8). It is thus hypothesized that this low-density anomaly is related to the Scari caldera (defined by Nappi et al., 1999) and its infill by highly vesiculated material.

Localized high-density anomalies found close to the coast agree well with the geology. The region of Piscità is covered by San Bartolo lava (red color in Fig. 1) and a corresponding high-density anomaly is found in this area (Figs. 5a and 7). Of our 8 samples at Stromboli (see sample 7 in Table 1 and Fig. 6), this lava features the highest saturated bulk density (2580 kg m$^{-3}$). High densities are likewise found in the areas of La Petrazza, Malpasso and Serro Barabba (Figs. 5a and 7); all corresponding to areas in which the oldest unit of Stromboli (Lower Paleostromboli I unit) is outcropping (darkest blue color in Fig. 1). These lava flows are dense (sample 1 in Fig. 6 has a saturated bulk density of 2450 kg m$^{-3}$).

*5.3. Time evolution of the density structure of Stromboli volcano*

Based on this first density model of Stromboli, an initial (somewhat speculative) proposition of the time sequence evolution in terms of density of the main geological objects is proposed in Figure 8:

1) During the period 85-41 ka (Paleostromboli I, II and Lower III), the island began its structuration in the present-day southern part of the island. Dyke intrusions began to mark the main structural directions of the island: N41 corresponding to the elongated shape of the island and N64, presenting important parallel structural incisions in the submarine northeastern part of the older Strombolicchio edifice (Romagnoli et al., 2009). Therefore, the first high-density anomalies could be related to this period.

2) During the period 41-34 ka (Upper Paleostromboli III and Scari unit), the low-density anomaly to the northeast could have appeared during this time due to the hydromagmatic Scari caldera formation.

3) During the period 26-4 ka (Vancori and Neostromboli), the main change in the density structure was caused by dyke intrusions. The end of this period was affected by the formation of the Neostromboli crater associated with the low-density anomaly in the summit area.

4) During the period 2-1.2 ka (Pizzo activity), the Neostromboli crater was refilled by vesiculated material (pyroclastites and scoriae).

5) In the period <1.2 ka (Present-day activity), dyke intrusions along N41 direction followed the eruptive center towards the present-day crater terrace. A strong density contrast resulted due to the main structural events that occurred in the summit area since the formation of the Neostromboli crater.

Our model of the density distribution of Stromboli volcano can be interpreted as a sum through time of endogenous constructive events (dyke intrusion) causing high-density anomalies and explosive (phreato-magmatic) destructive events causing low-density anomalies. The importance of the two main faults N41 and N64 is highlighted by dyke intrusions in the central and southern part of the island. In contrast, the northern part of the island does not seem to be affected by significant dyke intrusions (i.e., no large-scale high density anomaly is found in this region).

## 6. Conclusions

The pronounced topography at Stromboli, together with a detailed DEM of the aerial and submerged part of the volcano, made it possible to directly invert the 543 land-based and 327 sea-surface gravity data for a 3-D density model with corresponding error levels of 0.1 mGal and 5 mGal. Two prominent low-density anomalies correspond to (1) the area of the Pizzo ridge up to the Neostromboli crater to the east and (2) the possible location of the Scari caldera above San Vincenzo village. The Neostromboli crater is surrounded by dense bodies that are especially dense towards the southwest (Sciara del Fuoco, Vallone di Rina) and towards the east in the Rina Grande. The two main faults in term of higher gas permeability (N41 and N64; Finizola et al., 2002) seem to play a major role in the location of the higher density anomalies on the island. One of the future goals of our research is to develop a conceptual model of Stromboli and its functioning that explains all types of available geophysical data and is in agreement with present-day geological understanding.

**Acknowledgements**

We thank the Civil Protection COA personnel for the logistical support during the field campaigns. We are most grateful to Maria Marsella and Silvia Scifoni from University of Roma "La Sapienza" who provided the 1 × 1 m DEM model of Stromboli and to Claudia Romagnoli from University of Bologna, Daniele Casalbore from University Roma Tre, and Francesco Latino Chiocci from University of Roma "La Sapienza" who provided the 15 × 15 m bathymetric model. We also like to express our gratitude to Alessandro Bonforte and all GPS staff from INGV Osservatorio Etneo for providing GPS monitoring data that served as our base station for the DGPS. We are grateful to the Herbette Foundation for covering the field expenses. A. Revil was supported by DOE (Energy Efficiency and Renewable Energy Geothermal Technologies Program) grant awards #GO18195 and #DE-EE0005513. A. Finizola was supported by CNRS-INSU funding. Comments from Giovanna Berrino and an anonymous reviewer helped to improve the clarity of the presentation.

**Table and Figure Captions**

**Table 1**: Densities and porosities of the rock samples analyzed in this work (Figure 6).

| Sample # | Type | Grain density [kg m$^{-3}$] | Dry bulk density [kg m$^{-3}$] | Saturated bulk density [kg m$^{-3}$] | Porosity [%] |
|---|---|---|---|---|---|
| 1 | Paleo-Stromboli I (lava flow) | 2680 ± 10 | 2310 ± 10 | 2450 ± 30 | 13.8 ± 0.8 |
| 2 | Paleo-Stromboli I (explosive) | 2570 ± 10 | 1930 ± 30 | 2180 ± 50 | 24.9 ± 1.5 |
| 3 | Vancori | 2630 ± 10 | 2370 ± 10 | 2470 ± 20 | 9.7 ± 0.7 |
| 4 | Vancori altered | 2700 ± 10 | 1920 ± 10 | 2210 ± 30 | 29.0 ± 0.7 |
| 5 | Neostromboli (old + dense) | 2790 ± 10 | 2230 ± 10 | 2430 ± 20 | 20.3 ± 0.7 |
| 6 | Neostromboli (vesiculated) | 2770 ± 20 | 1610 ± 10 | 2030 ± 20 | 42.0 ± 0.6 |
| 7 | San Bartolo (lava flow) | 2830 ± 10 | 2440 ± 10 | 2580 ± 20 | 13.8 ± 0.6 |
| 8 | Rina Grande (reworked) | 2740 ± 20 | 1280 ± 20 | 1810 ± 50 | 53 ± 1.0 |

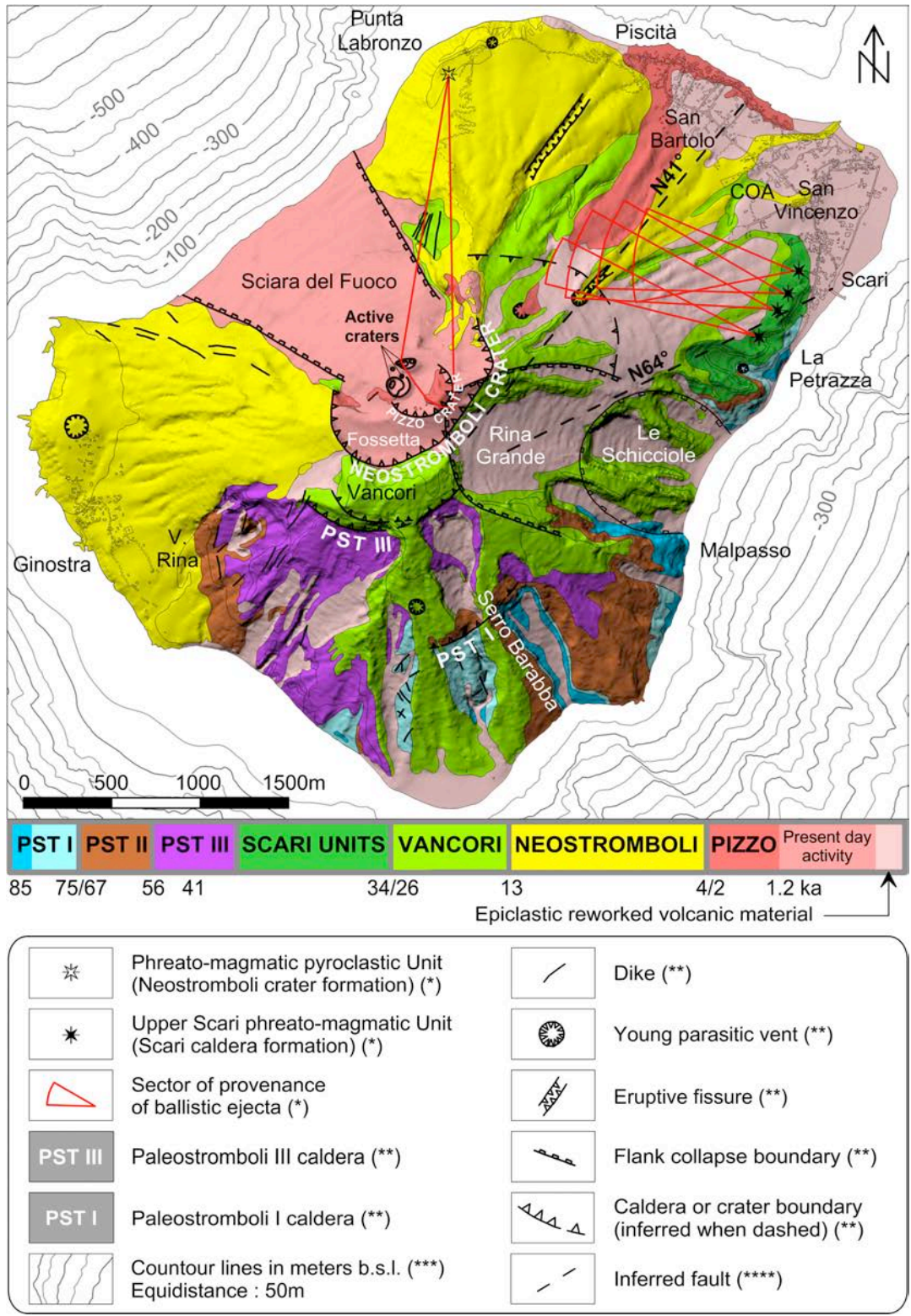

**Fig 1.** Simplified geological map of Stromboli volcano showing the different stages constituting the evolution of the edifice (modified from Keller et al.,1993 and Francalanci et al., 2013). (*): after Nappi et al., 1999; (**): after Keller et al., 1993; (***): after Romagnoli et al., 2009; (****): after Finizola et al., 2002.

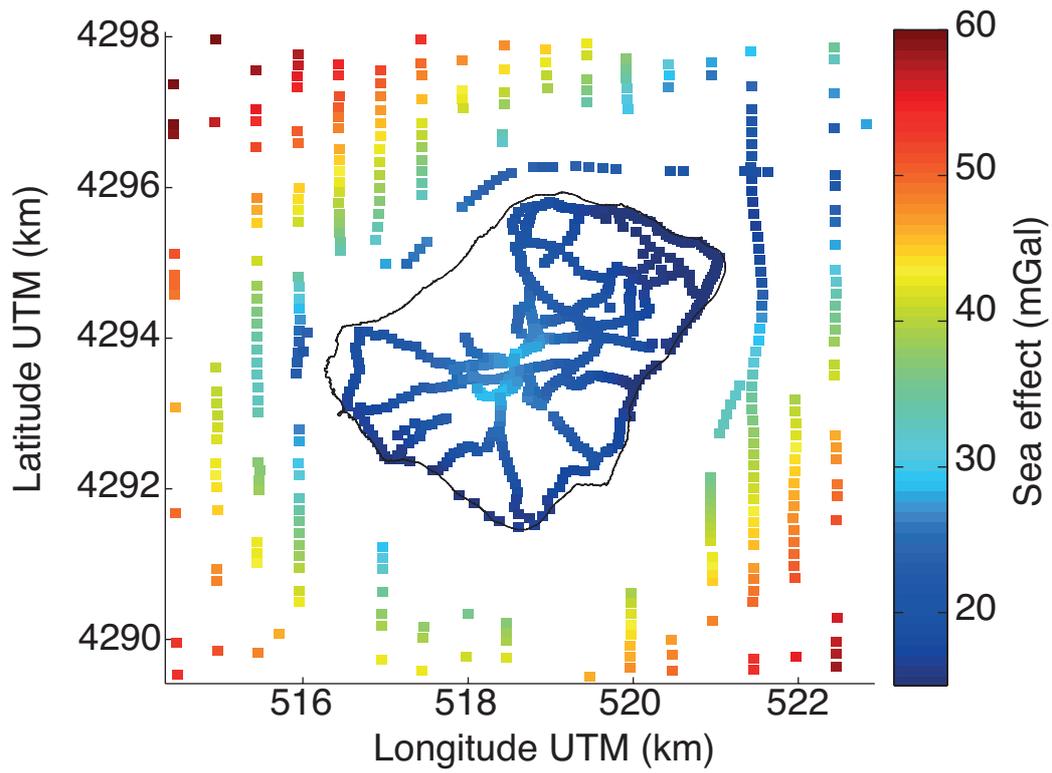

**Fig. 2.** Vertical component of the gravitational attraction exerted by the sea. The sea effect varies between 15 and 29 mGal for the land-based measurements and between 17 and 63 mGal for the sea-surface data.

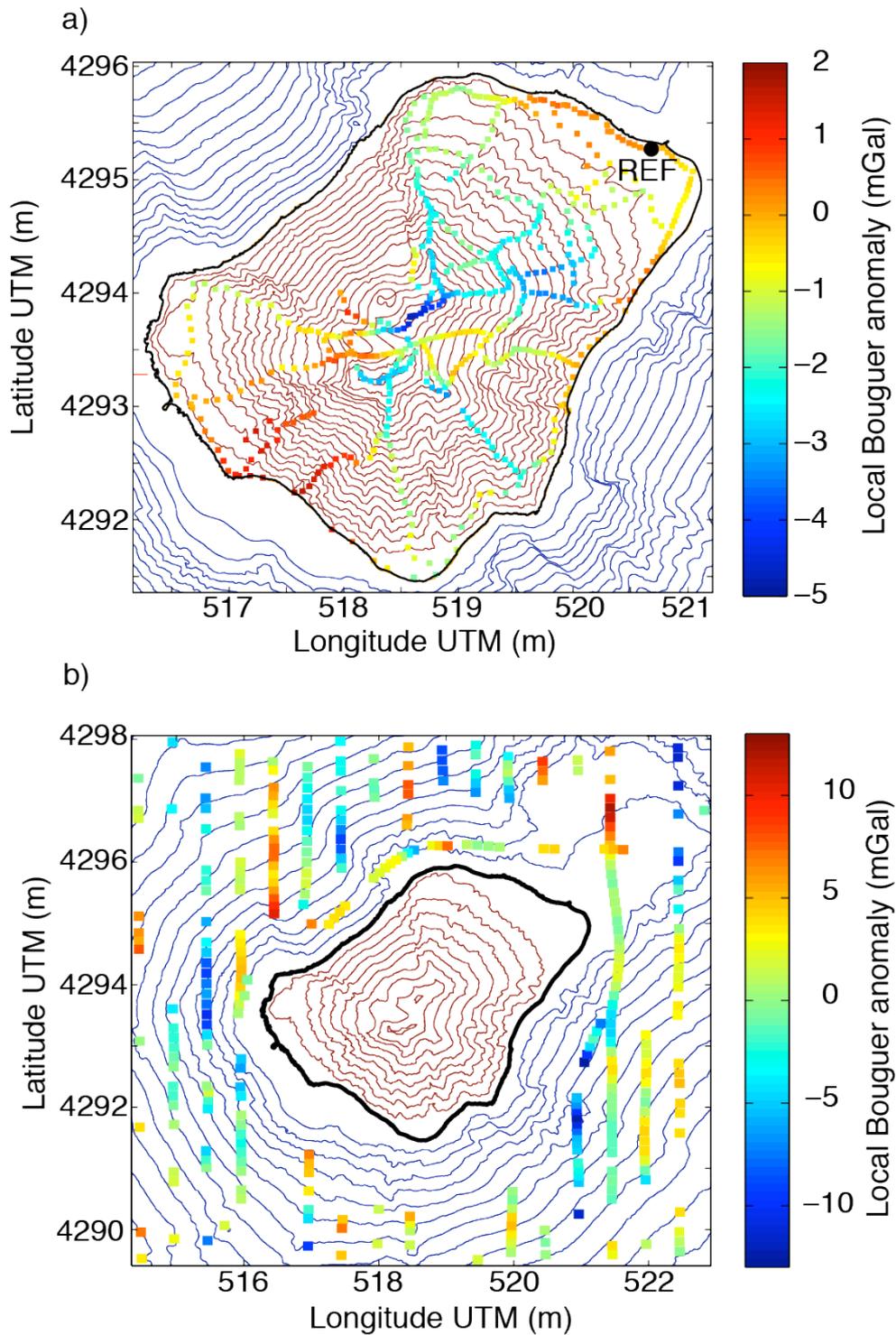

**Fig. 3.** (a) Local Bouguer anomaly of the (a) land-based (density 2100 kg m$^{-3}$) and (b) sea-surface (2600 kg m$^{-3}$) data. For the land-based data, the anomalies are given with respect to the position indicated as "REF" in (a), while the sea-surface data are given with respect to the mean of all sea-surface data. Note that the regional trend and the sea effect have been removed from this representation. The isolines outline the island's topography and the bathymetry.

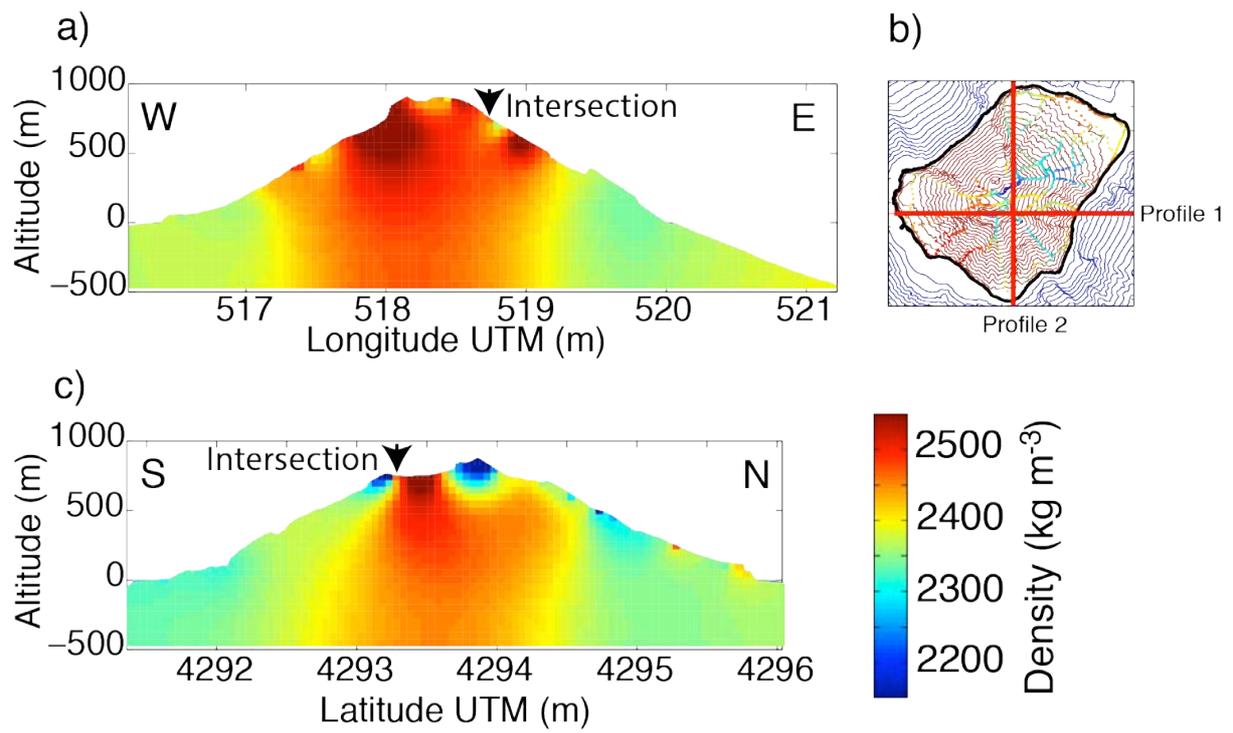

**Fig. 4.** Vertical slices of the 3-D density model. (a) W-E trending slice indicated as Profile 1 in (b) and (c) S-N trending slice indicated as Profile 2 in (b).

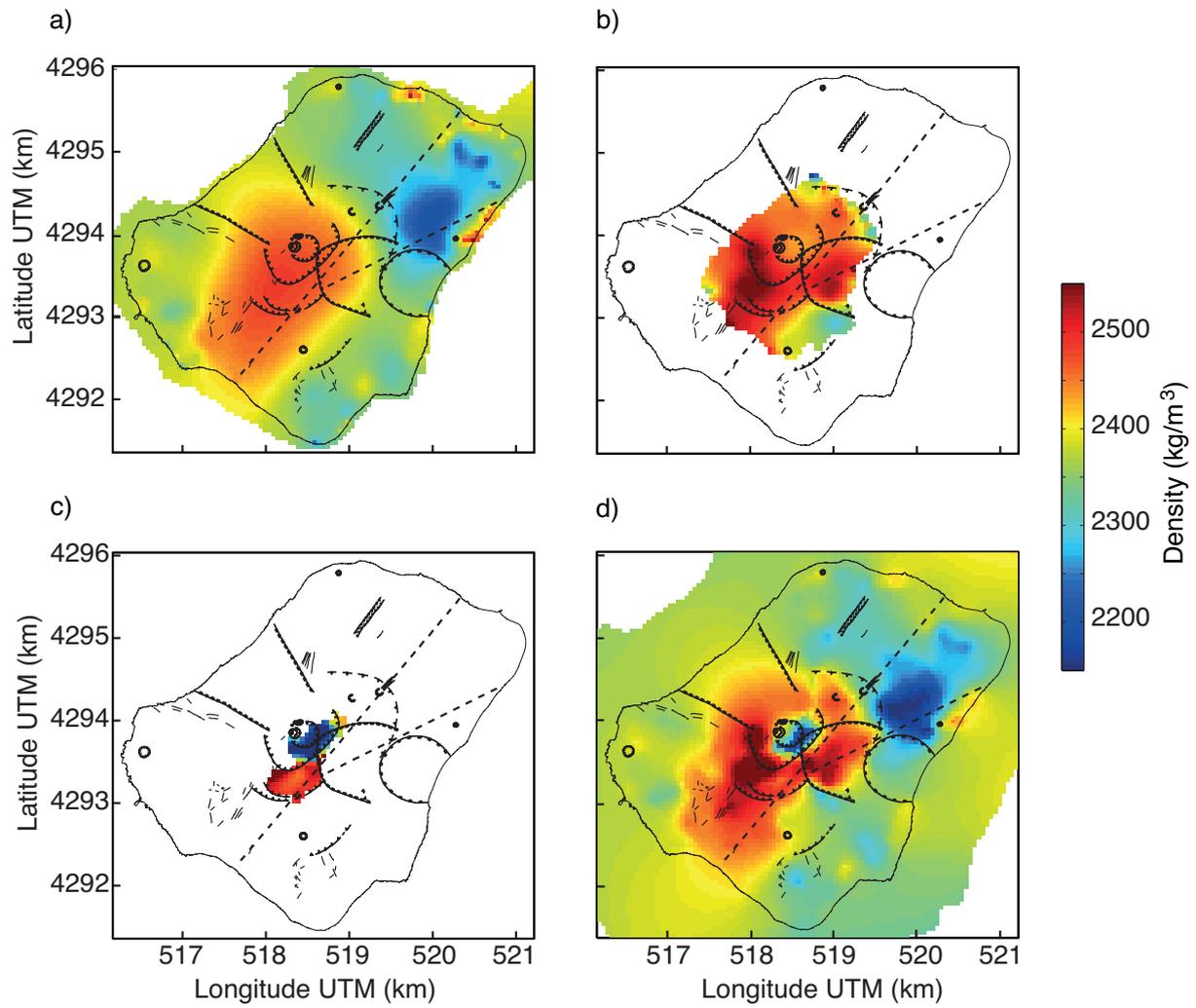

**Fig. 5.** Horizontal slices of the 3-D density model at (a) 0 m, (b) 500 m and (c) 800 m above sea level. (d) Slice at 125 m depth from the land surface (i.e., parallel to the topography). The main structural boundaries and the coastline of Stromboli are outlined.

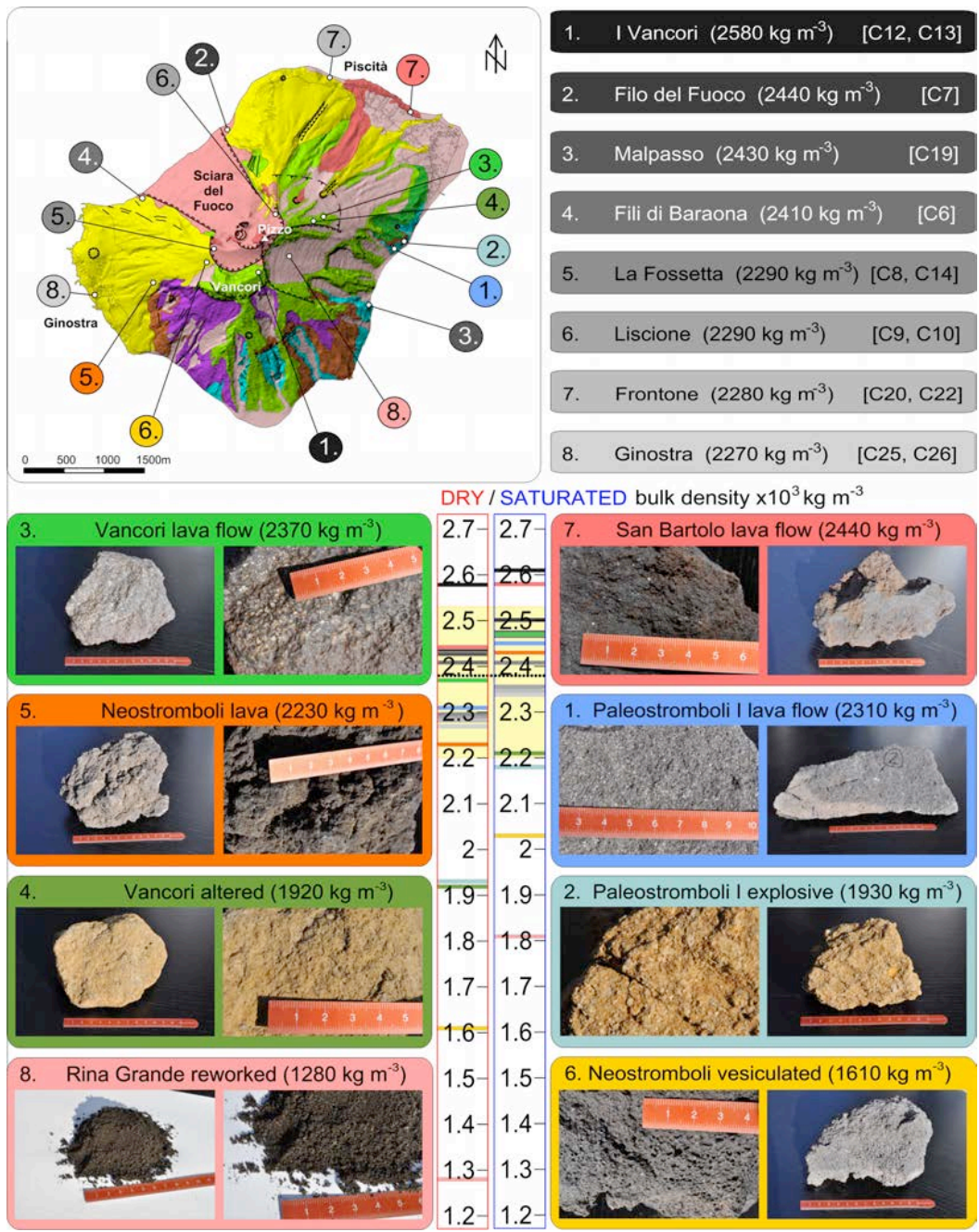

**Fig. 6.** Locations and pictures of our 8 samples collected on Stromboli with dry bulk density values (color background), and the samples location from Apuani et al. (2005) with dry bulk density values (black-grey background) with Cxx corresponding to the numeration of the samples in Apuani et al. (2005). All these values are represented on a density scale (middle of the figure) with the same color code for dry (on the left) and water saturated (on the right) bulk density. The black horizontal dotted line represents the mean density (2380 kg m$^{-3}$) obtained by inversion of the gravity data at a depth of 125 m from the land surface. The light yellow background rectangle displays the 2.5 and 97.5 percentiles of the density model obtained by inversion at this depth.

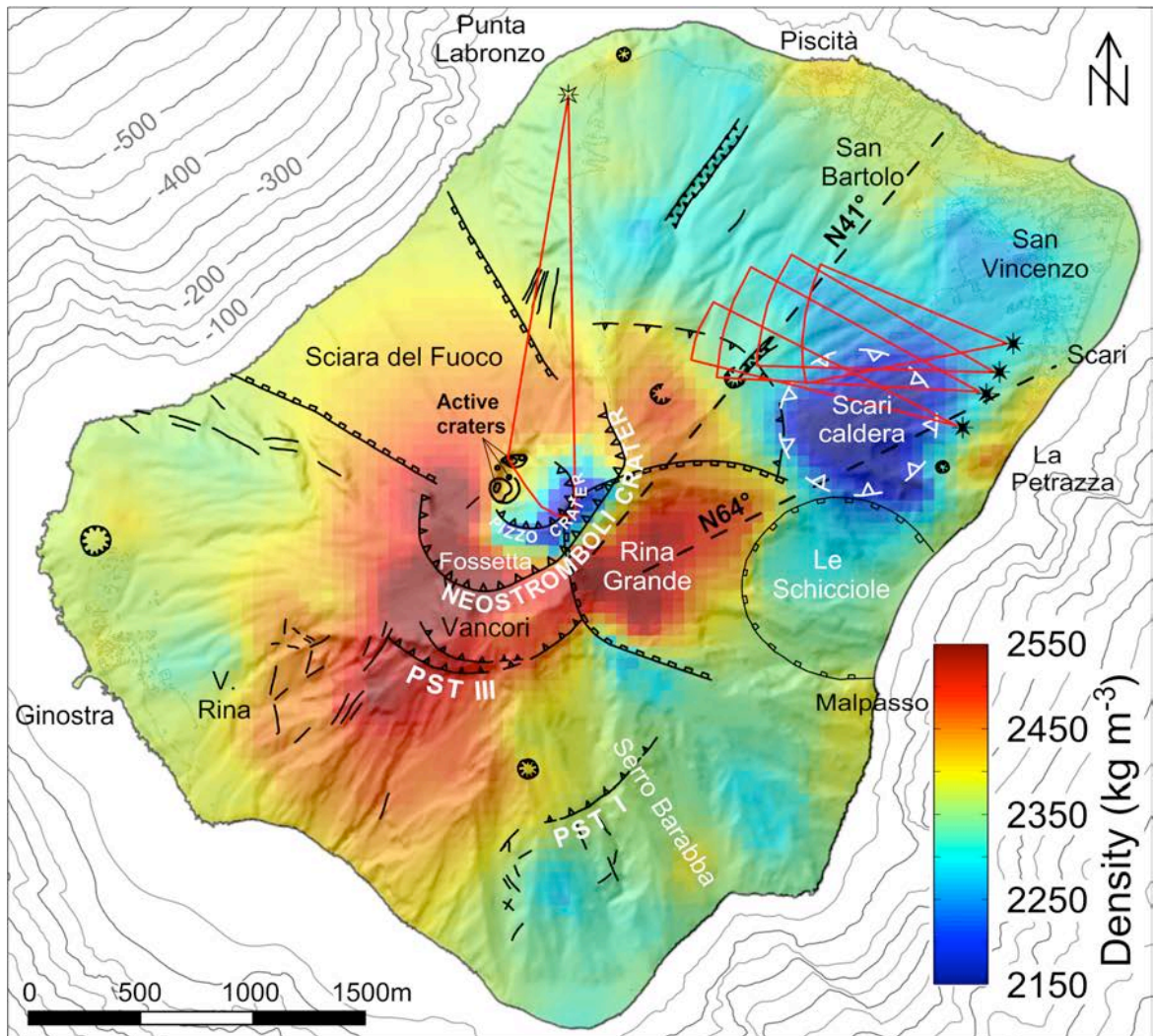

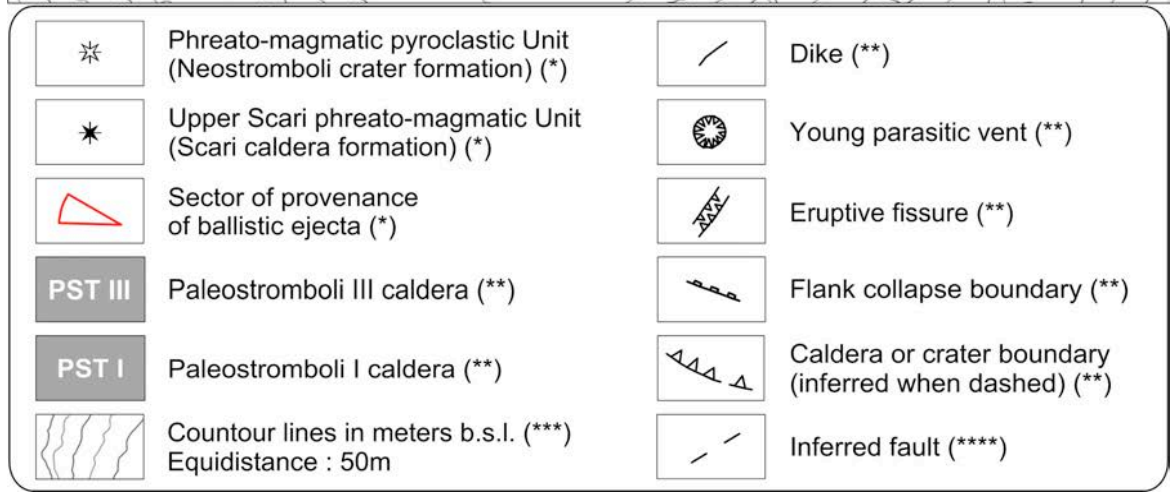

**Fig. 7.** Density model slice at 125 m depth from the land surface superimposed on the main structural boundaries of Stromboli island. The density model allows hypothesizing the location of the Scari caldera (white triangles). (*): after Nappi et al., 1999; (**): after Keller et al., 1993; (***): after Romagnoli et al., 2009; (****): after Finizola et al., 2002.

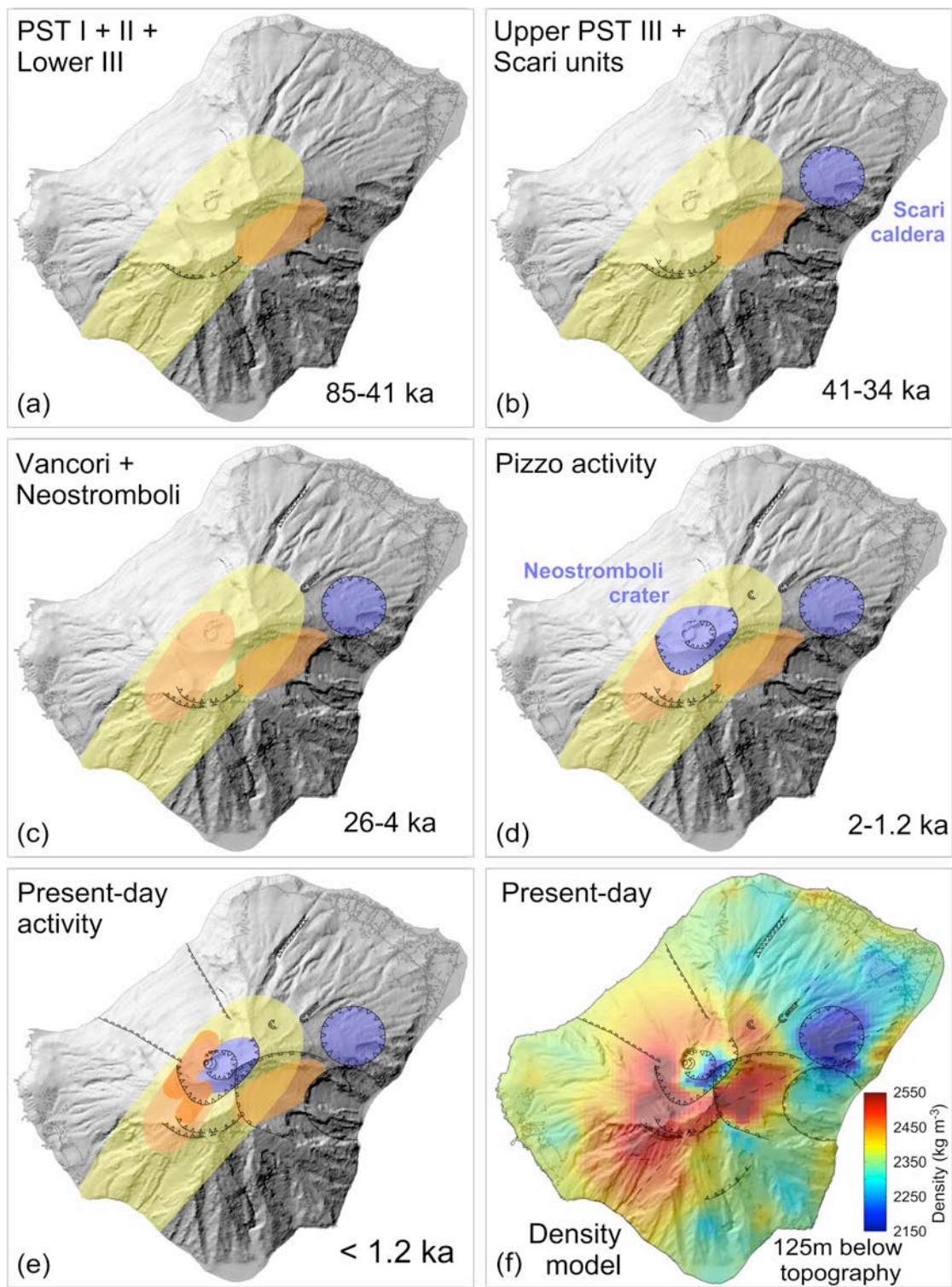

**Fig. 8.** Schematic interpretation of how the main dyke intrusions have, over time, created high-density regions (yellow-orange-red areas) and how the main phreato-magmatic explosions have created low-density anomalies by the deposition of vesicular material (dark blue areas).

31